\documentstyle[trans]{rspublic}
\input{epsf.sty}
\newcommand{\decsec}[2]{$#1\mbox{$''\mskip-7.6mu.\,$}#2$}
\begin{document}
\title{The Sloan Digital Sky Survey}

\author{Bruce Margon\thanks{On behalf of the entire scientific and technical 
team of the {\it Survey}, including personnel at Institute for Advanced Study,
Johns Hopkins University, Princeton University, University of Chicago,
University of Washington, Fermi National Accelerator Laboratory, the Japanese
Participation Group, U.~S. Naval Observatory, and several individuals at
external institutions.}}
\shortauthor{Bruce Margon}

\affiliation{University of Washington, Astronomy Department,\\
Box 351580, Seattle, WA 98195-1580, USA}

\maketitle
\label{firstpage}
\begin{abstract}

The Sloan Digital Sky Survey is an ambitious, multi-institutional
project to create a huge digital imaging and spectroscopic data bank of 25\%
of the celestial sphere, approximately 10,000 deg$^2$ centred on the north
galactic
polar cap. The photometric atlas will be in 5 specially-chosen colours, 
covering the $\pi$~ster of the Survey area to a limiting magnitude of
$r'$$\sim$23.1, on \decsec{0}{4} pixels, resulting in a 1~Tpixel map. This data
base will be automatically analysed to catalogue the photometric and astrometric
properties of $10^8$ stellar images, $10^8$ galaxies, and $10^6$
colour-selected QSO candidates; the galaxy data will in addition include
detailed morphological data. The photometric data are used to autonomously
and homogeneously select objects for the spectroscopic survey, which will
include spectra of $10^6$ galaxies, $10^5$ QSOs, and $10^5$
unusual stars. Although the project was originally motivated by the desire to 
study Large Scale Structure, we anticipate that these data will impact 
virtually every field of astronomy, from Earth-crossing asteroids to QSOs at 
$z>6$. In particular, the $\sim12$~TByte multi-colour, precision calibrated 
imaging archive should be a world resource for many decades of the 
next century.

\end{abstract}

\section{Introduction}

The Sloan Digital Sky Survey (SDSS) is arguably the most ambitious digital data
base of the celestial sphere yet undertaken. It employs a special purpose
2.5m telescope, an unusually large camera, and a pair of multi-fibre 
spectrographs, located on Apache Point, New Mexico.
Construction of all hardware and most of the software is now complete, and
observations of numerous primary and secondary standard stars to underlie the
specialised photometric system are underway. First light on the unique
mosaic camera, the most complex astronomical imaging instrument yet 
constructed, occurred in May 1998. After a commissioning period of 6--12
months, we hope to be obtaining `survey-quality' data, and it should require
5--6 years thereafter to complete the entire effort.

The title of the Survey commemorates the generous financial support of the 
Alfred P. Sloan Foundation, a philanthropic foundation located in New York. 
To date considerable additional financial support has been provided by 
the U. S. Department of Energy, the Japanese 
Participation Group, the member universities, the U. S. Naval 
Observatory, and the U.S.
National Science Foundation.

Participants in the construction and data-taking phases of the SDSS are from a
dozen institutions listed in the cover page footnote. The umbrella organisation
to execute this project is the Astrophysical Research Consortium (ARC), a 
non-profit corporation of 7 U.~S. educational institutions incorporated in 1984
to construct and operate astronomical facilities for its members. ARC also 
operates a 3.5m general purpose telescope on the same site at Apache Point. 
Although not all ARC member institutions participate in both SDSS and the 
3.5m projects, the synergy of the two telescopes is clear, as the 3.5m can 
spectroscopically reach essentially every object imaged by SDSS, while the SDSS
spectra cover only the brightest tip of the SDSS imaging archive.

Some excellent and only slightly dated review articles on SDSS have been 
presented by Gunn \& Knapp (1993), Gunn \& Weinberg (1995), and Fukugita 
(1998), and a summary suitable
for the general public may be found in Knapp (1997). Here I will concentrate 
on an overview of the concepts of the Survey and the unique contributions 
which we believe it will make to astronomy in general and Large Scale 
Structure (LSS)
in particular. However a recurring theme of this paper will be that LSS is 
only a small fraction of the total scientific import of SDSS.

\section{Concept and Approach}

\subsection{Scientific Goals}

As noted above, the SDSS is actually two surveys, one photometric and the other
spectroscopic. Although objects for the spectroscopic survey are autonomously 
and homogeneously selected from the photometric data, both of these data
bases serve multiple purposes. The imaging data base may broadly be said to
serve four functions:

$\bullet$~automated, homogeneous, and quantifiable target selection and 
astrometry for the
spectroscopic survey, based on colour and morphology of images that meet
pre-selected criteria for a multitude of different scientific projects

$\bullet$~{\it photometric} redshifts, accurate to $\Delta z\sim$.05, will be
available for $>10^7$ galaxies, by virtue of the carefully selected 5 colour
bands of the Survey

$\bullet$~a discovery catalogue of $10^6$ QSO candidates, virtually unbiased 
by redshift, and with high purity; $>60$\% of these objects should prove 
to be actual QSOs when verified spectroscopically, although we intend to 
obtain spectra of `only' $10^5$ of them

\medskip
\noindent
and, at least in my personal view, most important of all,

\medskip

$\bullet$~a permanent, public, well-calibrated archive of $10^8$ stellar
objects and $10^8$ galaxies, including 5-colour photometric, astrometric, and
morphological data for each entry

\medskip

The spectroscopic database is likewise multifunctional, including:

$\bullet$~$10^6$ homogeneous galaxy spectra from an exceptionally-well
characterised sample, for LSS as well as other studies

$\bullet$~$10^5$ homogeneous QSO spectra, which should span $0<z<7$ (if there 
are any objects at $z>6$!) with minimal bias in $z$

$\bullet$~$10^5$ unusual stellar objects, selected autonomously for
spectroscopy due to their abnormal position in 5-colour space with respect to
the normal stellar locus

\medskip

The imaging and spectroscopic observations are interleaved during the 5 year 
survey; at least in principle spectroscopy can occur in the next lunar cycle 
following the imaging of the target field.  This protocol takes best advantage
of changing sky and Moon conditions, especially given the disparate exposure
times for imaging {\it vs.} spectroscopy, and significantly 
shortens the total duration
of the Survey. It does however place severe demands on the complex software 
pipeline that acquires the data, performs image recognition, 
classification, astrometry, and precision, 
calibrated photometry, followed by target selection for a myriad of different 
scientific projects. This large system must work not only reliably but {\it 
rapidly} on a huge volume of data, if the insatiable demand for new
spectroscopic targets, sometimes as many as 6,000 per night, is to be
continually satisfied on only a few weeks of lead time.

\subsection{Hardware}

The hardware to complete the Survey is as unique as the scientific goals. The 
alt-az telescope is a special purpose, wide-field instrument, of 2.5m aperture,
designed and optimised specifically for this task, although much of its 
heritage may be traced back to the ARC 3.5m. Both instruments are located on 
Apache Point, New Mexico, a 2,780m peak located 2~km from Sunspot, New
Mexico, the site of the National Solar Observatory. Although the telescope is 
of conventional 
Cassegrain design, the 1.1m secondary mirror yields an unusual $3^\circ$ 
field of view at the f/5 focus where both imaging and spectroscopy are 
conducted; the plate scale there is $61\mu$~arcsec$^{-1}$. The large focal
plane is nearly a Mercator projection of the sky, and the optics deliver 
excellent images from $0.3-1.0\mu$ over this very large field.
The telescope, primary, and
secondary optics were fabricated by L\&F Industries, University of Arizona
Optical Sciences Center, and Steward Observatory Mirror Laboratory,
respectively, under contract to the University of Washington. 

The enclosure 
design by M3 Engineering is also unconventional, in that to minimise both 
equilibration times with 
the environment and building-induced seeing degradation, the 
small building which covers the telescope during the daytime and in inclement 
weather is retracted on rails during observations, leaving the telescope 
entirely exposed during data acquisition. The telescope is then protected from 
wind-shake by a baffle which cocoons it, and corotates but does not touch the
instrument. This fast (f/2.2 at primary) telescope is quite small and thus
stiff in any case. A second dedicated telescope of 0.5m aperture, used for
exhaustive absolute photometric calibration each night, is located nearby.

The heart of the hardware, and perhaps its most technically ambitious 
component, is the photometric camera (Gunn {\it et al}. 1998).
Despite recent and rapid advances in the technology of charge-coupled 
devices, it is a daunting task to cover the physically very large 
focal plane (650~mm) with
detectors; yet only with the large $3^\circ$ field can one hope to cover the
$10^4$~deg$^2$ desired for the Survey in a reasonable time interval. The
camera thus contains 30 CCDs each of $2048\times2048$ 24$\mu$ pixels,
yielding a respectable scale on the sky of \decsec{0}{4}~pixel$^{-1}$. At our
desired limiting magnitudes and high galactic latitudes of the Survey region
(chosen in part to avoid large uncertainties of extinction
corrections), source confusion is thus not a major problem. The CCDs,
tilted to best map out the large focal plane, are
housed in six separate and almost contiguous dewars, each containing five
chips, one for each of the five colour bands of the Survey. In normal Survey
operations, imaging data are acquired via TDI (Time Delay and Integrate) 
mode, with
charge continually clocked down the chips and read out at the same rate that
the sky drifts past the focal plane. The effective integration time on any 
given point on the celestial sphere is 55~sec, which yields reasonably faint 
limiting magnitudes ({\it e.g}., $r'=23$) with a 2.5m telescope and these
highly efficient detectors. An alternative `staring' mode to
obtain images would entail very substantial inefficiencies due to the dead 
times while the $>$10$^8$ pixels in the focal plane are read out. 
The camera also includes on its perimeter 22 additional $2048\times400$ pixel 
CCDs of lesser sensitivity, to observe transits of brighter stars without 
saturation, and thus provides excellent astrometric solutions for the fainter 
data. This remarkable instrument was designed and constructed at the Princeton
University Observatory, with the CCDs supplied by Scientific Imaging 
Technologies. In the wavelength regions where it is possible to use thinned, 
back-side illuminated chips with the most favourable coatings (3 of the 5
bands), the quantum efficiency of these excellent detectors is $\sim$80\%.

\begin{figure}[h]
\centering
\vspace*{5pt}
\parbox{\textwidth}{\epsfxsize=3.5in \epsfbox{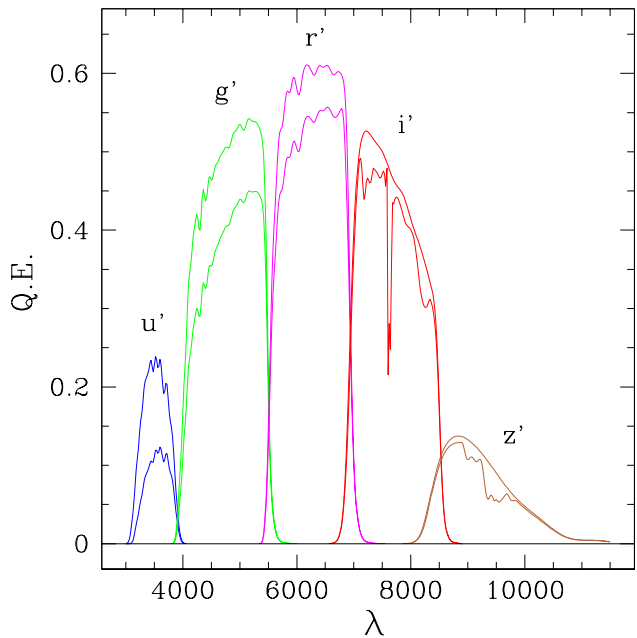}}
%\parbox{\textwidth}{\epsfxsize=\textwidth \epsfbox{fig1.ps}}
\vspace*{5pt}
\caption{The expected quantum efficiency of each Survey bandpass, including 
allowance for optics, filters, and detectors (Gunn {\it et al.} 1998). The
upper curve is `above the atmosphere,' and the lower includes 1.2 airmasses
of atmospheric extinction.
\label{fig:fig1}}
\end{figure}

The SDSS photometric system (Fukugita {\it et al.} 1996), comprised of five
largely disjoint bands centred from 3500--9500\,\AA, although derived from the 
Gunn-Thuan system, is also distinct to this project.
The Survey bandpasses are shown in figure~\ref{fig:fig1}, and are derived from
a number of constraints and compromises. The rationale to establish a new 
and unique photometric system should be compelling, and we believe that this
is indeed the case. The SDSS bands provide wide total coverage almost from
the UV-atmospheric cutoff to the near-IR silicon cutoff, are nearly uniformly
logarithmically spaced in wavelength, effectively exclude the $\lambda$5577
nightsky
emission, and are well-suited to surprisingly accurate determination of
photometric galaxy redshifts for a large range of $z$. The $u'$ band fits
nicely between
the atmospheric cutoff and the Balmer jump, the $g'$ and $r'$ bands are
reasonable matches to the $J$ and $F$ plates of the new Palomar Observatory
Sky Survey, and $i'$ ends just shortward of the bright OH nightsky emission.
The $z'$ band is well-suited 
for discovery of objects beyond the current distance record of $z$$\sim$5, 
as continuum absorption shortward of Ly$\alpha$ then suppresses images in 
other of the filters, and the very strong observed Ly$\alpha$ emission typical
of high-$z$ QSOs is concentrated in just one band.
Considerable early work, both theoretical (Newberg \& Yanny 1997, 
Lenz {\it et al.} 1998)
and observational (Richards {\it et al.} 1997, Newberg {\it et al.} 1997, 
Krisciunas {\it et al.} 1998) has 
already been invested in delineating the normal stellar locus, as well as the 
positions of a variety of interesting objects, in this photometric system, as 
this information is essential for efficient, autonomous selection of large 
numbers of unresolved images for many of the Survey's spectroscopic programs, 
as well as for transformation of SDSS magnitudes to other systems.

It was clear from the inception of the project that an ambitious multi-fibre 
spectrograph would be needed to reach the goal of $10^6$ homogeneous galaxy
spectra and $10^5$ QSOs. To the desired limiting magnitude for our LSS galaxy
survey, $r'\sim$18.2, each 7~deg$^2$ field contains 400--500 galaxies, and
thus our design has allowed for 640 fibres per field. It should be recalled
that the SDSS is not a general purpose telescope or observatory, and the
spectrograph is designed for one repeated, homogeneous purpose with the very
highest efficiency. Thus a basic philosophical decision was made not to
employ robotic positioning of fibres, as has been popular with many recent
multiobject spectrometers. We have instead opted for reasons of simplicity
and reliability for what might perhaps be viewed as a step backwards in
technology, namely a metal plate custom-drilled for each field by a
numerically-controlled milling machine. The hole positions are derived from
astrometric solutions of the SDSS imaging data, to position and retain 640
fibre optic cables. The $180\mu$ fibres subtend $3''$ on the sky, and may be
spaced by as little as $55''$ before mechanical interference becomes a
problem. Overlapping plates will be required to achieve the desired
homogeneity of the sample. The plates and fibres are packaged and positioned
by a rigid cartridge assembly which will be plugged by hand in the daytime
prior to observing each night; 9 identical assemblies insure that there are
ample targets to fill even a long, clear winter night, given 45 minutes for a 
typical exposure and 15 minutes to change cartridges. At the risk of stating 
the obvious, such a night yields 6,000 spectra! 
Bookkeeping of the correspondence of a given fibre to a given hole in the 
aperture plate is automated, by illuminating the fibres from the spectrograph 
slit end,
and observing the resulting pattern on the plate.

This design approach has the downside
of requiring procurement of a very large number of high quality fibres, 
hopefully of homogeneous properties and uniformly high throughput. 
We have indeed succeeded at this formidable requirement, having received and 
tested more than 6,000 fibres with a mean throughput of 92\%.
The system also includes dedicated fibres for guiding and sky monitoring.
The fibre system has been 
discussed in detail by Owen {\it et al.} (1998).

The spectrographs themselves consist of two identical modules, each receiving 
the light from 320 of the fibres. The fibres are spaced by $390\mu$ at the
slit entrance, slightly more than twice their diameter on the sky, so
bleeding and scattered light between adjacent fibres should be very minimal.
The spectra cover the 3900--9100\,\AA\ range with
resolution $\lambda/\Delta\lambda\sim$1800, or $\sim150$~km~s$^{-1}$, 
comparable to the velocity dispersion of a typical galaxy observed. 
Each spectrograph in turn has two 
channels, with the red and blue light split by a dichroic optic onto pairs of 
$2048\times$2048 CCDs with $24\mu$ pixels, identical to the chips in the 
mosaic camera; the CCD electronics for the spectrographs 
are also identical. These instruments,
designed and built at Johns Hopkins and Princeton, are very simple, with
virtually no moving parts, and with a grism as the dispersing element should
provide excellent throughput, $>30$\%.

\subsection{Data Handling and Software}

Even a casual reader can appreciate that the large number of pixels in the 
mosaic camera imply that the final SDSS data base must be very large, but some
specific values are instructive. The camera produces 8~MB~sec$^{-1}$ of 
data, which will total nearly 12~TB by the end of the $10^4$~deg$^2$ survey.
As the $10^6$ spectra are one-dimensional, that volume of raw data 
is `small,' about 
400~GB. The data are recorded on multiple tape drives, and sent each day from 
the Observatory by express air courier to Fermilab for routine processing and 
photometric calibration, using a large software package written primarily at 
Princeton University and Fermilab. An 
`operational' data base is used for automated target selection for the 
spectroscopic programmes, while a compressed, fully calibrated `science 
data base,' whose architecture was developed at Johns Hopkins, 
will be the focus of most of the analysis, both by the SDSS team
and also the general astronomical community. 
For example, almost all of the original information is 
retrievable from a set of `atlas images,' small regions of sky around every 
well-detected object (as well as interesting objects catalogued at other 
wavelengths but undetected in SDSS),
which total about 700~GB. While the raw data directly 
off of the CCDs will also be stored indefinitely, and in several locations, we
envisage that the atlas images, together with a merged pixel map of the entire
Survey area, will serve the vast majority of scientific
investigations. Given the rapid and continuing decrease in price for mass disk
storage, it seems plausible that by the time the Survey is complete, most 
large observatories and astronomy departments will opt to keep this version
of the data online at all times.

Many astronomers do not routinely work with data volumes of this size. One 
rapidly appreciates that even the simplest operations (`select every object in 
this portion of sky with these colours and that morphology') are not practical,
at least not with current and foreseeable desktop hardware, 
unless great care is devoted not just to the query algorithms, but to the 
structure of the data base. We will use a hierarchically-structured, 
object-oriented data base architecture, with {\it Objectivity}, a commercial 
software environment, as the basic underlying engine.

All SDSS data will be 
entirely public, on a schedule limited chiefly by the human resources 
needed to process and calibrate this very large volume of information.

\section{Survey Strategy}

\begin{figure}
\centering
\vspace*{5pt}
\parbox{\textwidth}{\epsfxsize=\textwidth \epsfbox{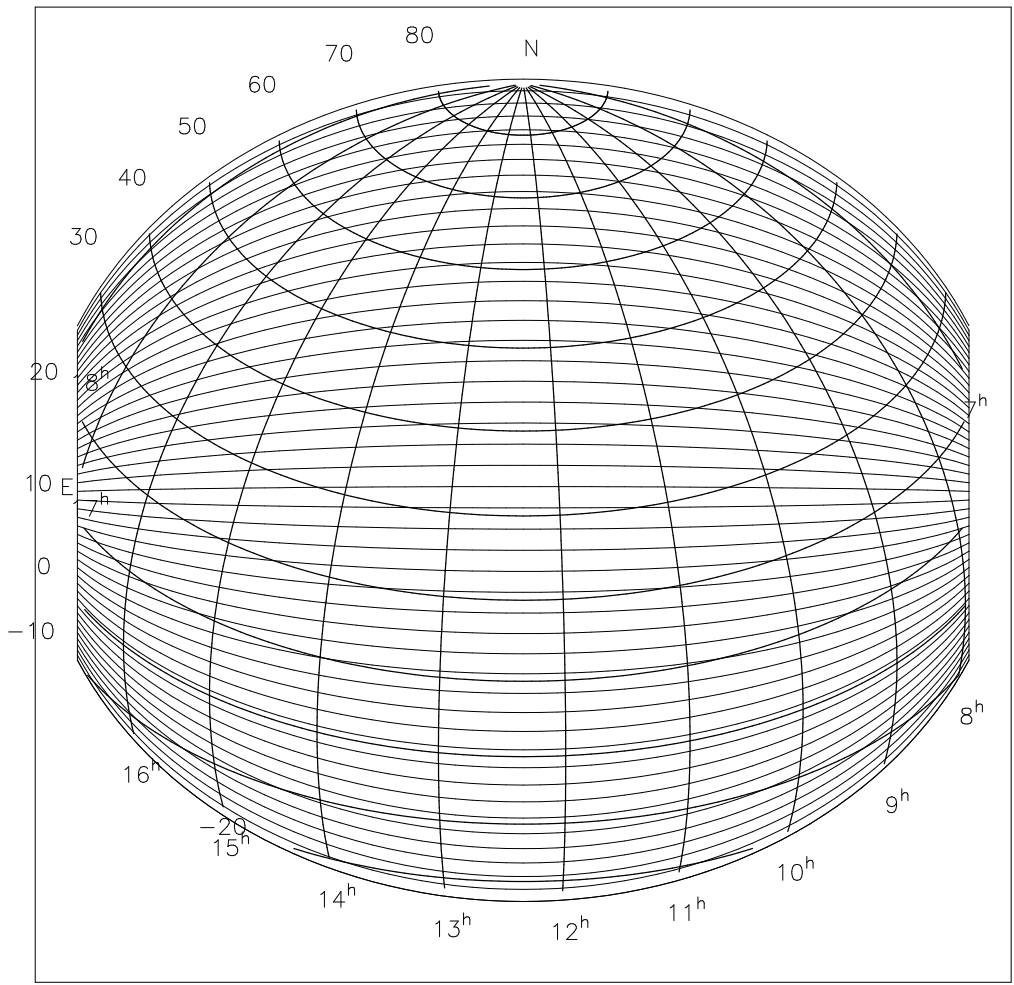}}
\vspace*{5pt}
\caption{The $10^4$~deg$^2$ survey region of the {\it Sloan Digital Sky 
Survey}.
\label{fig:fig2}}
\end{figure}

Even the seemingly straightforward task of selecting the $10^4$~deg$^2$ 
survey region is in fact 
subtle. A region centred precisely on the North Galactic Pole and confined to
$b>30^\circ$ does not best minimise galactic foreground extinction -- a very 
serious issue even at low extinction levels for any magnitude-limited 
extragalactic survey -- nor provide an optimal range of telescope 
zenith distances given 
our Observatory's latitude of $+33^\circ$. We have instead chosen a somewhat 
oblique region that spans $130^\circ$ E/W and $110^\circ$ N/S, centred at
$\alpha=12^h20^m, \delta=+32.5^\circ$. This region is displayed in 
figure~\ref{fig:fig2},  and
has also been chosen by the FIRST team for their 1400~MHz comprehensive radio 
survey made with the VLA in the B configuration (Becker {\it et al}. 1995).
Thus the entire SDSS volume will also have 1~mJy ($5\sigma$) radio coverage
with $\sim1''$ accuracy positions for point sources.

An original scientific motivation for SDSS was of course an intensive study of
Large Scale Structure, and the sample of $10^6$ galaxies will 
indeed dominate the spectroscopic portion of the project, requiring 400--500 
fibres of the 640 total on each plate. This large spectroscopic data bank
will reach $r'\sim$18.2, with $<$$z$$>$\,$\sim0.1$ and $z_{max}\sim0.25$. 
We stress however that it is not simply the size of this sample that is of 
interest, but rather the excellent level at which it is characterised. We will
know the morphology of each galaxy, and limiting areal magnitudes in five 
colours to a few hundredths accuracy, in a strictly homogeneous system. As 
selection bias is the bane of extragalactic astronomy, this virtue of the SDSS
sample cannot be overstated.

An ideal goal 
would be to probe the power spectrum on scales to 500~Mpc, yielding elegant 
overlap with COBE data, as well as anticipated second-generation microwave
background surveys. In practice, the effects of small and irregular
foreground extinction may be severe, causing spurious signals for scales of
250~Mpc or more. We will thus use the spectra of hot halo stars in each field,
together with galaxy counts and H\,I/IR data, to estimate the extinction in as
many independent ways as possible. We may or may not reach this 500~Mpc goal, 
but are confident that it will be difficult to do better by any other 
technique. The combination of results from large surveys such as SDSS and 2dF 
(Colless 1998)
with upcoming microwave background measurements from MAP and Planck will be 
particularly exciting, with prospects for remarkably precise and reasonably 
model-independent determinations of $H_0$ and $\Omega$ ({\it e.g.,} Eisenstein
{\it et al.} 1998, Gawiser \& Silk 1998).

About 100 fibres per spectroscopic plate will be devoted to a large QSO 
survey. Candidates for spectroscopy are, like the galaxies, selected 
autonomously from the SDSS photometric data, although in this case primarily
colours rather than morphology are employed. Simulations and some limited
actual data (Richards {\it et al}. 1997) indicate that the 5 SDSS photometric
bands are highly efficient at identifying QSO candidates at all redshifts;
only a relatively narrow band at $z\sim2.5$, where the QSO locus passes
through the normal stellar locus, is problematic. Furthermore, the 
spectroscopic integration times are optimised for galaxies, but clearly will 
yield useful results considerably fainter for QSOs, where all the light falls 
in the fibre, and strong emission lines permit easy redshift determination. We
thus expect SDSS to yield $10^5$ spectra of (essentially all 
previously uncatalogued) QSOs, to a limit of $g'\sim19.7$. This sample is 
$10\times$ larger than the sum of all QSOs currently known from all sources. 
Aside from the 
considerable value to QSO physics of such a homogeneous spectroscopic sample,
with precision colours and astrometry as well, we can anticipate 
$\sim3\times10^5$ absorption line systems in the data bank. While our spectral
resolution is less than ideal for such work, these data do provide an 
additional probe of LSS quite separate from the galaxy sample.
Finally, large systematic searches such as SDSS are surely 
the best way to identify the brightest 
members of any sample, in this case useful for later absorption line studies 
at large telescopes and from space.

Colour selection from the SDSS imaging data will provide $10^6$ additional
QSO {\it candidates} to $i'<21.5$ over the $\pi$~ster survey region. While we
have insufficient fibres to observe each, this will be a finding list of 
high purity ($>60$\%) for future work, and also contain some extremely 
high-$z$ objects.
Uncertain but 
reasonable extrapolations (Schneider 1998) imply that SDSS
may find $\sim40$ objects at $z>6$.

There will typically be several dozen fibres per spectroscopic plate remaining
after the LSS galaxy and QSO spectroscopic surveys receive their allocation,
even after additional fibres are used for various overheads such as sky
background, guiding, photometricity monitoring, etc. These excess fibres will
be used for a wide variety of scientific programs in stellar and
extragalactic astronomy, probably limited chiefly by the imagination of our
team.  The majority of them will be allocated autonomously via colour and
morphological selection from the imaging data base. (Although the capability 
does exist to manually specify an individual target for spectroscopic 
observation, SDSS is hardly a good venue for such work; one might just as 
easily walk 20 metres down the road and use the ARC 3.5m general purpose 
telescope.) We will for example take spectra of as many as possible of objects
that lie far off the 5-colour locus of normal stars, but for some reason have
not been selected for the QSO survey, as well as objects which coincide with 
FIRST radio and ROSAT All Sky Survey X-ray positions.  We thus expect to 
discover numerous unusual stars in the halo, ranging from distant carbon stars
and RR~Lyrae's (invaluable as halo dynamic probes) to cataclysmic variables
and other degenerate binaries, extremely metal poor stars, planetary nebulae,
dwarf carbon stars, hot white dwarfs, etc. etc.

In the autumn months the Survey region is not accessible at our observing site,
but a reasonable amount of high galactic sky near the south polar cap is 
available. There is little virtue to repeating the primary Survey protocol 
during this time, as the total Survey volume is not then rapidly increased.
Instead we plan to perform a separate deep imaging program which we refer to 
as the `Southern Survey' (not however to be confused with the southern 
hemisphere). Over the 5 years of the main survey, we will repeatedly image a 
225~deg$^2$ stripe in this region, on approximately 45 separate visits. The 
sum of these data will go $\sim2$~mag fainter than the main Survey, {\it e.g.}
to $r'=25.1$, as well as probe yet another new portion of discovery space, 
namely the nature of time variability of the faint high latitude sky on a 
timescale of $\sim1$~month. These data also provide an empirical determination
of the completeness of many aspects of the main Survey. Various
special-purpose spectroscopic programmes may also be performed here.

\section{Scientific Strengths of the SDSS}

In a project that performs such a comprehensive survey, together with a 
large and disparate number of targeted
investigations, ranging from Earth-crossing asteroids to $z>6$ QSOs, it is 
most certainly a matter of personal taste to rank order, or even briefly 
but completely enumerate, the scientific strengths of the programme. Nonetheless
it seems an appropriate summary to at least attempt to do so. Therefore I list
here a strictly personal view, in inverse order (from least to most 
important), what I view as the scientific strengths of SDSS.

{\it Large data base}: The five-year Survey will produce $10^6$ galaxy
spectra, $10^5$ QSO spectra, $10^6$ QSO images, $10^8$ galaxy images with
$>10^7$ photometric $z$'s, and $10^8$ stars with 5-colour photometry.

{\it Homogeneity of the data:} The Survey yields
$10^6$ galaxies and $10^5$ QSOs in the identical spectrograph with very high 
signal-to-noise; additionally
$10^8$ galaxies and $10^8$ stellar images are catalogued 
with 5-colour, very well-standardised
photometry and excellent astrometric accuracy.

{\it Exceptionally well-characterised samples:}
Each of the galaxies selected for the LSS survey has quantitative 
morphological data, and extremely accurate,
5-colour photometry. At the risk of stating something very 
well-known to all LSS pundits, {\it LSS conclusions are known to depend
on colour and morphology!} Numerous recent authors {\it e.g.}, Loveday 
{\it et al.} 1995, Hermit {\it et al.} 1996, Guzzo {\it et al.} 1997) stress 
that one's knowledge of the precise composition of the galaxy sample will
heavily influence one's conclusions. Even such prosaic issues as slightly 
inaccurate
removal of minor, patchy amounts of foreground extinction can play havoc with
a magnitude-limited sample. Similar considerations apply to QSOs; in SDSS
they are selected from 5-colour space, with only one narrow deadband near
$z\sim2.5$. Of course, the impact of galaxy morphology
on LSS is not just an annoying selection effect, but something
interesting to study in its own right; it yields important inferences
on the bias of galaxies relative to dark matter, and therefore
on galaxy formation. 

{\it Science not related to LSS:}
~For stellar population studies, we have $10^6$ galaxy spectra with the
identical spectrograph, {\it with detailed morphology and 5-colour photometry
available for each object.} Large numbers of previously-unrecognised 
clusters of galaxies, low surface brightness galaxies, and gravitational 
lenses will be 
identified. The Survey produces $>10^5$ unusual stellar
spectra;
there are implications not
just for stellar astronomy, but for galactic structure, high-latitude
extinction, etc. Numerous ($\sim10^5$) asteroids will be found and 
characterised photometrically, including not 
just main belt objects, but also Near Earth Objects, Centaurs, Kuiper Belt 
objects, and long period comets.

{\it Discovery potential:}
Many spectroscopic fibres are available in each field, unrelated to the
LSS or QSO surveys, for objects of odd colour and/or morphology, FIRST radio
sources, {\it ROSAT} All Sky Survey X-ray sources, etc. The filter system is
especially well-optimised for $z>5$ QSOs.

{\it Archival value of the imaging data bank:}
The 5-colour, \decsec{0}{4} pixel, extremely well-calibrated imaging
archive of the entire northern sky at $b>30^\circ$ will be a
community resource for decades.
This ambitious but we believe correct statement seems the most appropriate 
with which to conclude this review.

\end{document}